\documentclass{PoS}

\def\src{PKS\,1510--089}
\def\fermi{\textit{Fermi}-LAT}

\usepackage[rightcaption]{sidecap}
\title{MAGIC observations of variable very-high-energy gamma-ray emission from PKS1510-089 during May 2015 outburst}

\ShortTitle{MAGIC observations of PKS1510-089 in May 2015}

\author{\speaker{Julian Sitarek}$^{,a}$, Josefa Becerra Gonz\'alez$^{b,c}$, Vandad Fallah Ramazani$^{d}$, Elina Lindfors$^{d}$, Giovanna Pedaletti$^{e}$, Fabrizio Tavecchio$^{f}$, Monica Vazquez Acosta$^{b,c}$, Stefan Larsson$^{g}$ for the MAGIC and Fermi-LAT Collaborations, Kiran Baliyan$^{h}$, Navpreet Kaur$^{h,i}$, Sameer$^{h,j}$, Svetlana Jorstad$^{k,l}$, Claudia Raiteri$^{m}$\\
        a) University of \L\'od\'z, PL-90236 Lodz, Poland (E-mail: \email{jsitarek@uni.lodz.pl})\\
        b) Inst. de Astrof\'isica de Canarias, E-38200 La Laguna, Tenerife, Spain\\
        c) Universidad de La Laguna, Dpto. Astrof\'isica, E-38206 La Laguna, Tenerife, Spain\\
        d) Tuorla Observatory, University of Turku and Astronomy Division, University of Oulu, Finland\\
        e) Deutsches Elektronen-Synchrotron (DESY), D-15738 Zeuthen, Germany\\
        f) INAF National Institute for Astrophysics, I-00136 Rome, Italy\\
        g) KTH Royal Institute of Technology, Department of Physics and Oskar Klein Centre for Cosmoparticle Physics, AlbaNova, SE-10691 Stockholm, Sweden\\
        h) Physical Research Laboratory, Ahmedabad 380009, Gujrat, India\\
        i) Indian Institute of Technology, Gandhinagar 382355, Gujrat, India\\
        j) Department of Astronomy and Astrophysics, The Pennsylvania State University, 532-D, Davey Laboratory, University Park, PA 16802, USA\\
        k) IAR, Boston University, 725 Commonwealth Ave, Boston, 02215, USA; \\
        l) St.Petersburg State University, Universitetsky prospekt, 28, St. Petersburg, 198504, Russia\\
        m) INAF, Osservatorio Astrofisico di Torino, via Osservatorio 20, I-10025 Pino Torinese, Italy
}

\abstract{
PKS1510-089 is a flat spectrum radio quasar located at a redshift of 0.36. 
It is one of only a few such sources detected in very-high-energy (VHE, >100 GeV) gamma rays. 
Though PKS1510-089 is highly variable at GeV energies, until recently no variability has been observed in the VHE band. 
In 2015 May PKS1510-089 showed a high state in optical and in the GeV range. 
A VHE gamma-ray flare was detected with MAGIC at that time, showing the first instance of VHE gamma-ray flux variability on the time scale of days in this source. 
We will present the MAGIC results from this observation, discuss their temporal and spectral properties in the multi-wavelength context and present modelling of such emission in the external Compton scenario.}

\FullConference{35th International Cosmic Ray Conference --- ICRC2017-\\
		10--20 July, 2017\\
		Bexco, Busan, Korea}

\begin{document}
\section{Introduction}
\src\ is a bright flat spectrum radio quasar (FSRQ) located at a redshift of $z=0.36$ \cite{ta96}.
The source is one of only six blazars firmly classified as a FSRQ from which gamma-ray emission has been detected in the very-high-energy (VHE, $>100$\,GeV) range \cite{ab13}. 
The GeV gamma-ray emission of \src\ is strongly variable with the doubling time of flares as short as 1\,h \cite{sa13}.
Until 2015, the source was detected only twice in the VHE gamma ray band, both during long periods of enhanced optical and GeV gamma-ray activity \cite{ab13, al14}.
Interestingly, no variability could be claimed from those detections. 

Since 2013, the Major Atmospheric Gamma Imaging Cherenkov (MAGIC) telescopes are performing regular monitoring of \src .
In May 2015, a strong flare of \src\ was observed in GeV gamma rays by the Large Area Telescope (LAT) on board the \textit{Fermi} satellite, accompanied by high activity in the optical and IR bands. 
The high state triggered further MAGIC observations, which led to the detection of an enhanced VHE gamma-ray activity from the source.
We report on the observations of \src\ during the May 2015 flare, discussed in more detail in \cite{ah16a}.

\section{Instruments and data analysis}
During the May 2015 outburst \src\ was observed by multiple instruments in a broad range of frequencies from radio up to VHE gamma rays. 

VHE gamma-ray data were collected using the MAGIC telescopes.
MAGIC is a system of two 17\,m diameter, imaging atmospheric Cherenkov telescopes located on La Palma, Canary Islands \cite{al16a}.
The MAGIC telescopes observed \src\ for 5.4 hours between MJD 57160--57166.
The data were analyzed using MARS, the standard analysis package of MAGIC \cite{za13, al16b} using an additional LIDAR-based correction for the atmospheric transmission \cite{fg15}.
The source has been observed in GeV range by \fermi\ during its all-sky monitoring program \cite{Atwood09}. 
Details of the data analysis are described in \cite{ah16a}. 

X-ray observations were performed with X-ray Telescope (\textit{XRT}) \cite{2004SPIE.5165..201B} on board the \textit{Swift} satellite. 
During the period from MJD 57153 to 57167 the source was observed 16 times for total time of 26.6\,ks. 
Details of the X-ray data analysis are described in \cite{ah16a}. 
The state of the source in optical-UV range was monitored by the Ultraviolet/Optical Telescope (UVOT, \cite{po08}),  also on board the \textit{Swift} satellite.
Those data were analyzed following the method described in \cite{ra10}.

The source was also monitored in the optical R range by a 35\,cm Celestron telescope attached to the KVA (Kunglinga Vetenskapsakademi) telescope located at La Palma. 
The analysis of those data was performed as described in \cite{ni17}. 
The optical polarization observations were performed with a number of instruments: Nordic Optical Telescope (NOT), Steward Observatory, Perkins Telescopes, RINGO3, AZT-8, and LX-200 (see \cite{ah16a} for details).
We also use infrared observations obtained with SMARTS, TCS and MIRO (see \cite{ah16a} for details). 

\src\ is also monitored at 37 GHz frequency with Mets\"ahovi Radio Telescope  (see \cite{te98}). 
High resolution radio images of the \src\ jet were obtained at the frequency of 43~GHz with the Very Long Baseline Array (VLBA).
The VLBA data were reduced following \cite{jo05}. 

\section{Results}
\begin{SCfigure}[][p]
\centering
\includegraphics[width=0.55\textwidth]{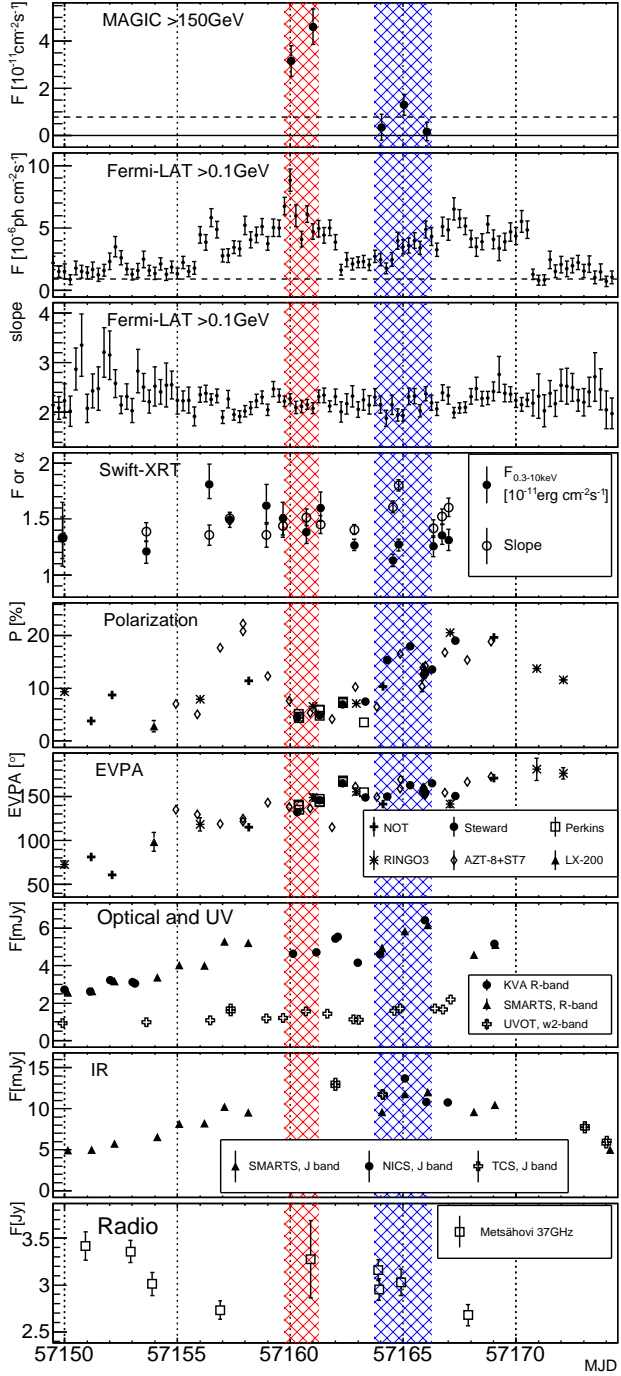}
\caption{
Multiwavelength light curve of \src\ during the May 2015 flare. 
From top to bottom: 
Nightly gamma-ray flux above 150\,GeV from MAGIC (the dashed line shows the average emission in Feb-Apr 2012, \cite{al14});
\fermi\ flux above 0.1\,GeV in 6\,h binning, and the corresponding spectral index (the dashed line shows the average emission from the 3FGL catalog, \cite{Acero15});
X-ray spectral flux (filled circles) and spectral index (empty circles) measured by \textit{Swift}-XRT;
polarization percentage and polarization angle measured by NOT, Steward, Perkins, RINGO3, AZT-8, and LX-200 (see legend);
optical emission in R band (KVA, SMARTS) and UV emission in w2-band (\textit{Swift}-UVOT);
IR emission in J band (SMARTS, MIRO-NICS, TCS);
radio observations by Mets\"ahovi at 37\,GHz. 
Data from IR  to UV are corrected for  Galactic absorption.
The red and blue shaded regions show  Period A and Period B, respectively, for which the spectral modeling is performed.
Figure reprinted from \cite{ah16a}.
}\label{fig:mwllc}
\end{SCfigure}

In Fig.~\ref{fig:mwllc} we show the multiwavelength light curve of \src\ during the period of MJD 57151--57174.
The VHE gamma-ray flux measured by MAGIC shows clear variability, with a chance probability of constant flux of just $7.7\times10^{-8}$. 
During MJD 57160--57161 the VHE gamma-ray flux was $\sim$ 5 times  higher than detected during 2012 \cite{al14}.
Afterwards (MJD 57164--57166) the source returned to the flux level compatible with the 2012 detection. 
Following those two states of the source we define two periods: A and B respectively, in which multiwavelength SED is investigated.
Despite the difference in the flux level, the spectral shape measured by MAGIC in both periods is consistent with each other and with previous measurements, however statistical uncertainties are rather large (see \cite{ah16a} for details). 

The GeV gamma-ray flux of \src\ measured by \fermi\ is highly variable in the whole investigated period (MJD 57150-57175). 
A few individual flares are visible, with time scales of a few days. 
A major GeV flare from \src\ also occured $\sim$60 days after Period A (see \cite{ah16a}). 
We reconstructed GeV spectrum in periods A and B. 
Similarly to the VHE gamma-ray case, while the flux level is different, no significant change of shape is observed. 

The X-ray flux, measured by \textit{Swift}-XRT, shows a gradual decrease during the period MJD 57156--57165. 
The X-ray emission also became significantly softer between Period A and B. 

The optical emission of \src\ during the period MJD 57150--57175 shows variability, which however does not strictly follow the gamma-ray one.
Similar behaviour is also seen in IR range. 

Throughout the investigated period, a smooth rotation of optical EVPA by $\sim 100^\circ$ occurred. 
The rotations of optical polarization angle has also been observed in the 2009 and 2012 gamma-ray flaring states \cite{ma10,al14}.
Nevertheless, the rotations of the EVPA are a common phenomenon in \src , therefore, further data are needed to firmly associate them with the emission of VHE gamma rays.
The low percentage of polarization, observed also during Period A, is typical for this source \cite{je16}.
The percentage of the polarization is three times higher both during Period B, and also a few days before Period A. 
The polarization rotation during the 2015 flaring period agrees with what is expected from a knot following a spiral path through a mainly toroidal magnetic field \cite{ma10}. 
Alternatively, it can be also  explained by the light travel time effects within an axisymmetric emission region pervaded by a predominately helical magnetic field \cite{zh15}.

The radio flux of \src\ shows moderate variability in the Mets\"ahovi observations performed at 37\,GHz.
It does not show any clear correlation with other bands, however, the sampling is rather sparse.

\begin{figure}[t]
\centering
\includegraphics[width=0.98\textwidth]{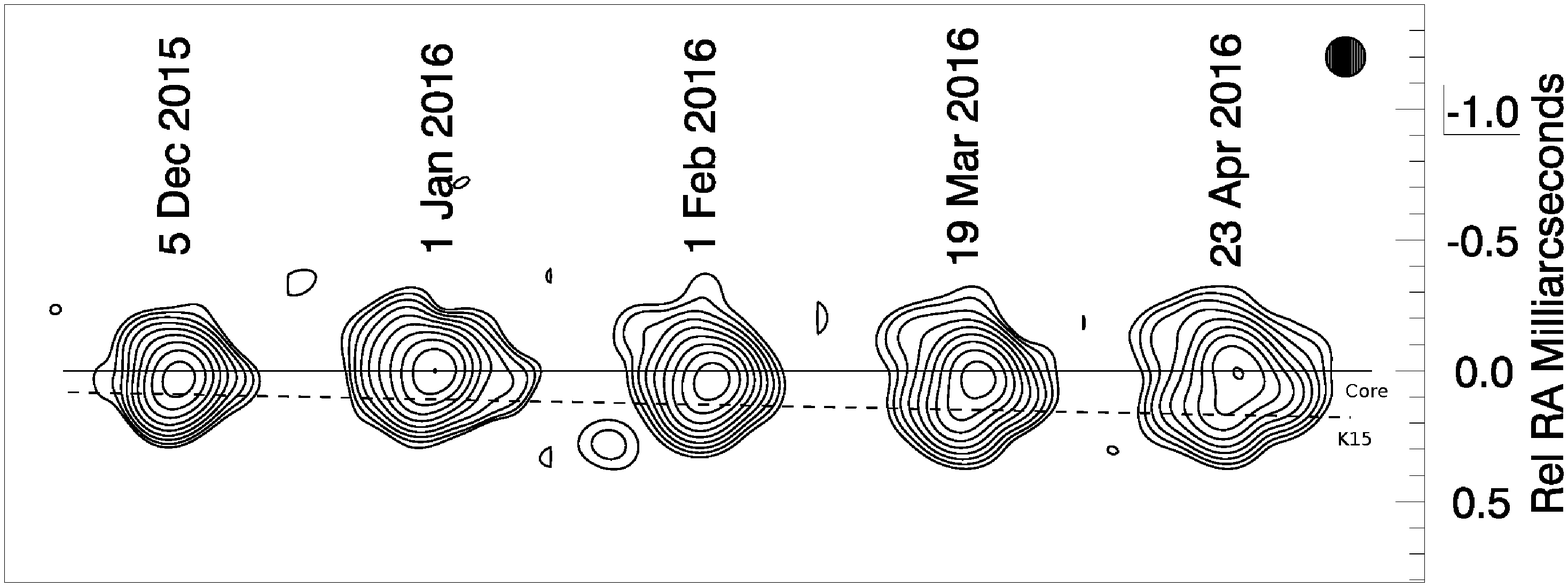}
\caption{
Total intensity images of the PKS1510-089 core region at 43 GHz, with a global peak intensity of $I_{peak}=3.566$\,Jy/beam and 0.15~mas FWHM circular Gaussian restoring beam (top right circle). 
The solid and dashed lines follow the positions of the VLBI core and $K15$, respectively, across the epochs.
Figure reproduced from \cite{ah16a}.
}\label{fig:1510t}
\end{figure}

VLBA observations of \src\ performed a few months after the flaring period revealed an occurrence of a new knot, K15, emerging from the core (see Fig.~\ref{fig:1510t}).
The new knot is bright and relatively slow, with an apparent speed $\beta_{app}$=(5.3$\pm$1.4)~c. 
Extrapolation shows that its separation from the core happened on MJD $T_0=57230\pm52$.
A similar behavior has also been observed during a high gamma-ray state in Feb-Apr 2012, when the emergence of a new radio knot, K12, from the core was associated with a VHE outburst \cite{al14}.
K15 core separation epoch is marginally consistent with the time during which MAGIC has observed VHE gamma-ray emission from \src . 
It should be noted however that due to large uncertainty of $T_0$ it could be also associated with one of a few GeV flares in this period (see \cite{ah16a}). 

\section{SED modeling}
The gamma-ray emission of FSRQs is typically explained in terms of the inverse Compton scattering of electrons on a radiation field external to the jet (see, e.g. \cite{sbr94,gh10}), the so-called external Compton (EC) scenario.
The type of the radiation field is determined by the location of the emission region.
The observation of VHE gamma rays escaping from the emission region suggests that the emission region is located outside the Broad Line Region (BLR) (see also \cite{ab13,al14}). 

\begin{SCfigure}[][t]
\includegraphics[width=0.55\textwidth]{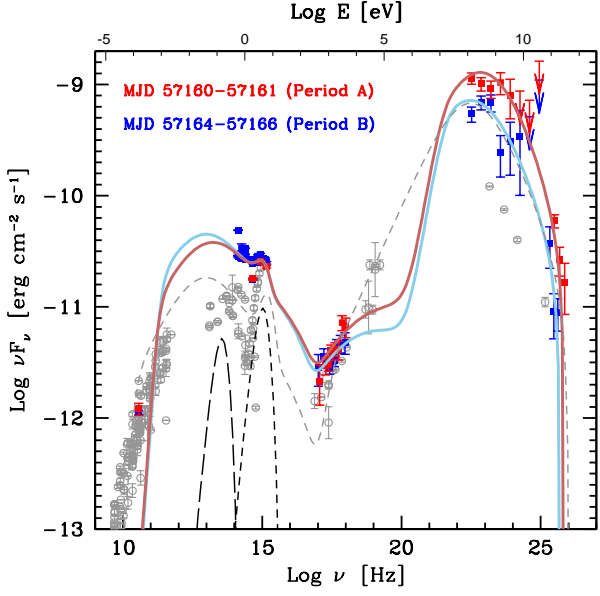}
\caption{
Multiwavelength spectral energy distribution of \src\ in Periods A (red symbols) and B (blue symbols). 
The red and the blue curves show the result of the emission model for the two periods. 
The black dashed and long-dashed lines show the adopted emission for the accretion disk and the dusty torus, respectively. 
For comparison, the dashed gray line shows the model derived for the SED in 2012 \cite{al14}.
Historical measurements (ASDC, see \texttt{http://www.asdc.asi.it/}) are shown as gray points.
Figure reproduced from \cite{ah16a}.
}\label{fig:sed}
\end{SCfigure}

In Fig.~\ref{fig:sed} we present the SED of \src\ constructed from the data covering Periods A and B, corresponding to high and low gamma-ray flux, respectively. 
Most of the flux variation (by $\approx$ a factor of 2--3) is visible in GeV and sub-TeV bands. 
The low-energy flux (optical, X-rays) is almost constant between the two periods. 
It is interesting to note that the high-energy peak during the period B is at a very similar level to the 2012 high state \cite{al14}, despite the IR--UV emission being a factor of $\sim3$ higher.

We model these SEDs of \src\ in the framework of a one-zone EC scenario, like the one used for the explanation of 2012 data  \cite{al14}.
In order to allow escape of VHE emission (observed by MAGIC) we assume that the emission region is located beyond the BLR radius.
Therefore, the external photon field seen by relativistic electrons is dominated by the thermal IR radiation of the dust torus (DT). 

To estimate the size and the radiation field of the BLR and DT we assume the scaling laws and the prescriptions given in \cite{gt09}. 
Assuming the disk luminosity of $L_{\rm disk}=6.7\times 10^{45}$ erg s$^{-1}$ \cite{al14} we obtain BLR and DT radii of $R_{\rm BLR}=2.6\times 10^{17}$ cm and $R_{\rm IR}=6.5\times 10^{18}$ cm respectively.
In calculations we assume that fractions $f_{\rm BLR}=0.1$ and $f_{\rm IR}=0.6$) of the disk radiation are intercepted and reprocessed by the BLR and by DT respectively. 
The DT is heated to 1000\,K. 

We fix the distance of the emission region from the base of the jet to $r=6\times 10^{17}$ cm.
If the emission region is filling the whole cross section of the jet, for an assumed jet semi-aperture angle $\theta_{\rm j}=0.047$\,rad we obtain the radius of the emission region $R=2.8\times 10^{16}$ cm.
Such a size of the emission region is consistent, even for moderate values of the Doppler factor, with the  variability observed by MAGIC with the time scale of a few days. 
We apply in the modeling the same values of the jet bulk Lorentz factor $\Gamma=20$ and Doppler factor $\delta=25$ as used in \cite{al14}. 
The remaining free parameters of the model are the intensity of the magnetic field $B$ and the electron energy distribution. 
Hence, we model the observed  variability as the effect of the changes in the conditions of the plasma flowing through the shock region.
To reproduce the SEDs  we assume that the electron energy distribution can be described by a double broken power law. 
The first break, $\gamma_c$, is caused by the cooling, and a second break, $\gamma_b$,  can be an effect of the acceleration process (see \cite{ah16a} for details). 
The used model can describe the data relatively well. 
The difference in the broadband emission of Period A and B can be explained with a relatively small change in the fit parameters, namely a slightly stronger magnetic field and lower maximum and break energies of the electrons during Period B. 

\section{Discussion and conclusions}
The observations performed by the  MAGIC telescopes revealed enhanced VHE gamma-ray emission from the direction of \src\ during the high optical and GeV state of the source in May 2015, showing for the first time VHE gamma-ray variability in this source. 
During May 2015 the IR, optical and  UV data showed a gradual increase in  flux, while the flux in the X-ray range was slowly decreasing. 

The May 2015 multiwavelength data are another example of  the enhanced VHE gamma-ray emission occurring during the rotation of the optical polarization angle. 
Also, similarly to other gamma-ray flares, an ejection of a new radio component was observed, however with a large uncertainty on the zero separation epoch, which makes it difficult to associate it to a particular peak in the GeV LC. 
Hence, May 2015 data suggests that the association of VHE gamma-ray emission with the rotation of EVPA and ejection of a new radio component might be a common feature of \src .

The source was modeled with the external Compton scenario. 
The evolution of the state of the source from the VHE gamma-ray flare to a weaker emission at the level of the 2012 detection can be explained by relatively small changes in the conditions of the plasma flowing through the emission region. 

Other scenarios might be also able to explain the observed emission. 
In particular, if we assume that the VHE flaring is indeed connected to the ejection of the new component (in the case of 2015 flare, $K15$) from the VLBA core and the rotation of the optical polarization angle, it would be natural to assume a single emission region located far outside the dusty torus.
The seed photons for EC process could then originate from the slower sheath of the jet. 
Such a scenario has been shown to provide a feasible description of the previous flaring epochs of \src\ (see \cite{al14,md15}). 

The VHE gamma-ray variability with time scale $\tau$ seen during the 2015 outburst puts constraints on the size, and therefore also on the location of the emission region. 
Assuming that the spine of the jet fills a significant fraction of the jet (as in \cite{al14}), the location of the emission region cannot be farther than $d=\tau \delta c / \left( (1+z)\theta_{\rm j}\right) = 2.7 (\tau/3\,\mathrm{days})(\delta/25)(\theta_{\rm j}/ 1^\circ)^{-1}$\,pc.
Therefore, a high Doppler factor and a narrow jet would allow us to place the emission region at the radio core. 
Such low values of the jet extension, $(0.2\pm0.2)^\circ$ \cite{jo05} and $0.9^\circ$ \cite{pu09} at the radio core are reported by the radio observations.
Intranight variability observed during the 2016 flare \cite{za17} will put even stronger constraints on the size and thus also location of the emission region.  
  
\section*{Acknowledgements}
We would like to thank the IAC for the excellent working conditions at the ORM in La Palma. We acknowledge the financial support of the German BMBF, DFG and MPG, the Italian INFN and INAF, the Swiss National Fund SNF, the European ERDF, the Spanish MINECO, the Japanese JSPS and MEXT, the Croatian CSF, and the Polish Narodowe Centrum Nauki. 
The \textit{Fermi}-LAT Collaboration acknowledges support for LAT development, operation and data analysis from NASA and DOE (United States), CEA/Irfu and IN2P3/CNRS (France), ASI and INFN (Italy), MEXT, KEK, and JAXA (Japan), and the K.A.~Wallenberg Foundation, the Swedish Research Council and the National Space Board (Sweden). Science analysis support in the operations phase from INAF (Italy) and CNES (France) is also gratefully acknowledged. This work performed in part under DOE Contract DE-AC02-76SF00515.

\end{document}